\documentclass[12pt,draftclsnofoot,onecolumn]{IEEEtran}

\usepackage{ifpdf}
\ifpdf
  \usepackage[pdftex]{graphicx}
  \usepackage[off]{auto-pst-pdf}
  \usepackage{subfigure}
  \graphicspath{{./pdfs/}{./eps/}{./jpg/}}
  \DeclareGraphicsExtensions{.pdf,.eps,.jpg}
  \input glyphtounicode\pdfgentounicode=1
\else
    \usepackage[dvips]{graphicx}
    \graphicspath{{./eps/}{./pdfs/}{./jpg/}}
    \DeclareGraphicsExtensions{.eps,.pdf,.jpg}
    \usepackage[cleanup={log,aux,dvi,ps},crop=on]{auto-pst-pdf}
    \usepackage{pst-pdf}
    \usepackage{subfigure}
\fi

\usepackage{color}
\usepackage{amsmath,amssymb,wasysym}
\usepackage{dsfont,pdfpages,euscript}
\usepackage{pdfsync}
\usepackage{caption}
\usepackage{rotating}

\allowdisplaybreaks

\begin{document}

\title{Interference Alignment Schemes Using Latin Square for $K \times 3$ MIMO X Channel}

\author{
Young-Sik Moon, Jong-Yoon Yoon, Jong-Seon No, and Sang-Hyo Kim
\thanks{Y.-S. Moon, Y.-J Yoon, and J.-S. No are with the Department of ECE, INMC, Seoul
National University, Seoul 88026, Korea (e-mail: myskill@ccl.ac.kr, yoon@ccl.snu.ac.kr, jsno@snu.ac.kr).}
\thanks{Sang-Hyo Kim is with the Department of ICE, Sungkyunkwan University, Suwon, 16410, Korea (e-mail: iamshkim@skku.edu).}
}

\maketitle

\begin{abstract}
In this paper, we study an interference alignment (IA) scheme with finite time extension and beamformer selection method with low computational complexity for X channel. An IA scheme with a chain structure by the Latin square is proposed for $K \times 3$ multiple-input multiple-output (MIMO) X channel. Since the proposed scheme can have a larger set of possible beamformers than the conventional schemes, its performance is improved by the efficient beamformer selection for a given channel. Also, we propose a condition number (CN) based beamformer selection method with low computational complexity and its performance improvement is numerically verified.  
\end{abstract}

\vspace{5pt}
\vspace{5pt}

\section{Introduction}
Interference alignment (IA) is an important technique to manage interference in the wireless communication networks. To resolve the interference problem, an interference alignment scheme was recently proposed and has become a subject of special interest in the area of wireless communications. Cadambe and Jafar \cite{cadambe08} showed that each user in a multi-user interference channel can utilize half of all the network resources, which corresponds to achieving the maximum degrees of freedom (DoF). The key idea of this result is the IA, which maximizes the overlap of all interference signal spaces at each receiver so that the dimension of the interference-free space for the desired signals is maximized.\\ 
\indent The idea of IA was further developed by many researchers [2]-[4]. 
In \cite{xchannel}, an IA scheme for $M \times N$ X channel was proposed and it was proved that the maximum DoF of the $M \times N$ X channel is $MN/(M+N-1)$. However, an infinite time extension is required to achieve this maximum DoF. In \cite{b1}, an IA scheme for $K \times K$ X channel was proposed without channel extension, where the beamforming vectors were constructed only by a spatial signature over unit time.\\ 
\indent Although most of the studies for IA focus on network throughput, that is, DoF, reliability is also an important performance measure of wireless communication systems. For reliability, various selection schemes have been studied such as transmit antenna selection or equivalent path selection based on singular value decomposition (SVD). Since signal to noise ratio (SNR) at the receiver is the most critical factor, most of the selection schemes for IA are based on SNR at receiver. However, obtaining the received SNR at the transmitter for beamformer design is practically difficult due to the high computational complexity.\\ 
\indent In this paper, we study an interference alignment (IA) scheme with finite time extension and beamformer selection method with low computational complexity for X channel. An IA scheme with a chain structure by the Latin square is proposed for $K \times 3$ multiple-input multiple-output (MIMO) X channel. Since the proposed scheme can have a larger set of possible beamformers than the conventional schemes, its performance is improved by the efficient beamformer selection for a given channel. Also, we propose a condition number (CN) based beamformer selection method with low computational complexity and its performance improvement is numerically verified.\\The rest of this paper is organized as follows: In Section II, the system model of $K\times 3$ MIMO X channel is described. Then, an expanded beamformer set using a Latin square for $K \times 3$ MIMO X channel is proposed in Section III. A couple of efficient beamformer selection methods are proposed in Section IV and its performance is numerically analyzed in Section V. Finally, conclusion is given in Section VI.
\vspace{5pt}

\section{System Model}
\begin{figure}[t!]
\centering
\includegraphics[scale=0.36]{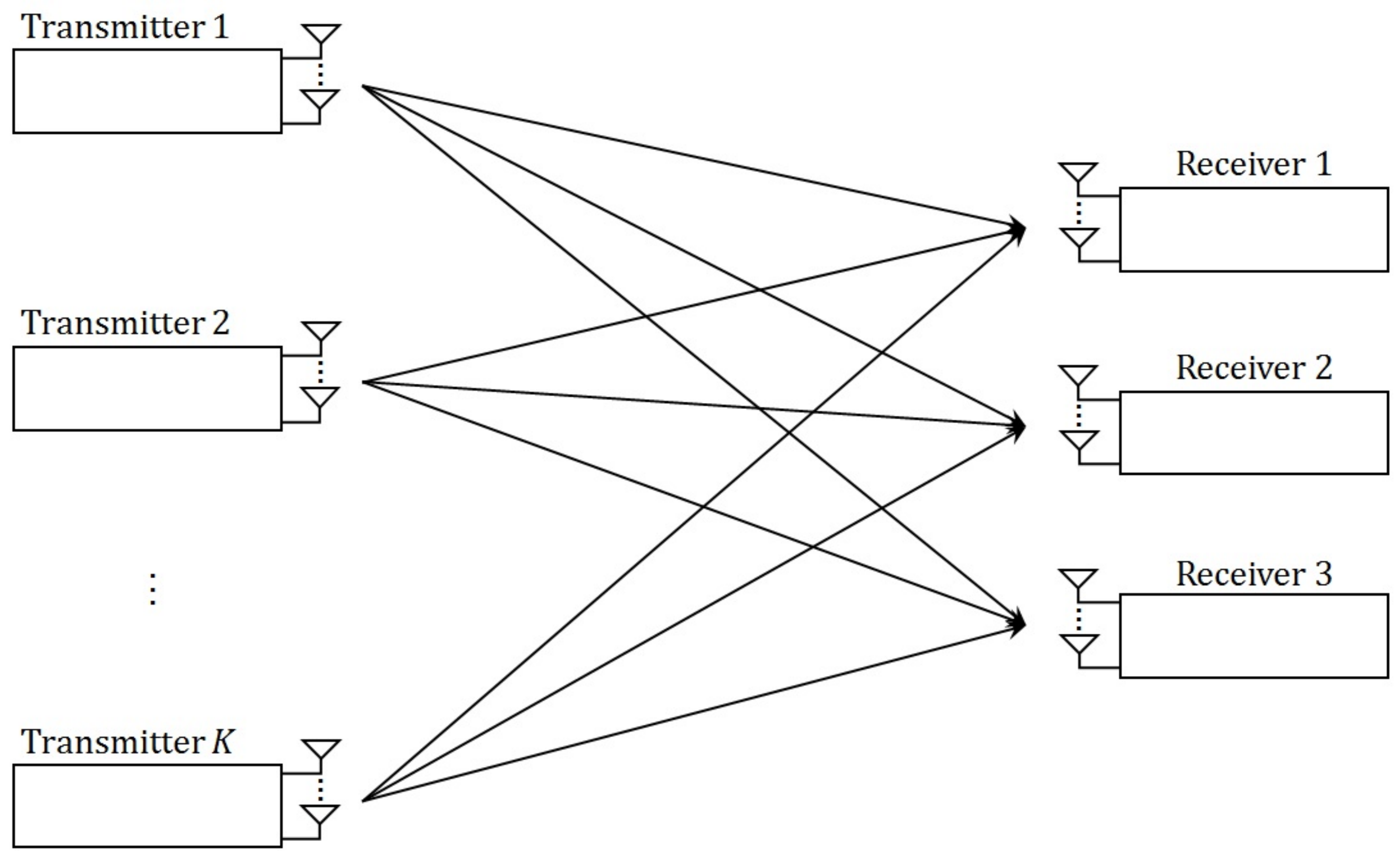}
\caption{$K\times 3$ MIMO X channel.}
\label{fig_pre_MIMO}
\end{figure}
We consider $K\times 3$ MIMO X channel as Fig. 1, where each transmitter $T_j$, $j=1,...,K,$ transmits independent message to each receiver $R_i$, $i=1,2,3$. All the nodes are equipped with $M=2K$ antennas.
The $2K \times 1$ transmit signal vector $\textbf{x}_j$ from the $j$-th transmitter can be represented as a linear combination of three different beamforming vectors
\begin{align}
\textbf{x}_{j}=\sum_{i=1}^3 \textbf{v}_{ij}s_{ij},     
\end{align}
where $s_{ij}$ denotes a transmitted message from the $j$-th transmitter to the $i$-th receiver and $\textbf{v}_{ij}$ denotes a $2K \times 1$ beamforming vector for the message $s_{ij}$.
The transmit signal vector $\textbf{x}_{j}$ has an average power constraint of $\mathrm{Tr}(\textbf{x}_{j}^{H}\textbf{x}_{j})\leq \text{P}_{j}$, where $\text{P}_{j}$ is the total transmit power of the $j$-th transmitter, $\mathrm{Tr}(\cdot)$ denotes the trace function, and $(\cdot)^H$ indicates the Hermitian transpose.\\
\indent Then, the received signal vector at the $i$-th receiver is given as
\begin{align}
\textbf{Y}_{i}=\sum_{j=1}^{K}\textbf{H}_{ij}\textbf{v}_{ij}s_{ij}+\sum_{j=1}^{K}\sum_{\substack{k=1 \\ k\neq i}}^{3}\textbf{H}_{ij}\textbf{v}_{kj}s_{kj}+\textbf{n}_{i},
\end{align}	
where $\textbf{H}_{ij}$ is the $2K \times 2K$ channel matrix from the $j$-th transmitter to the $i$-th receiver generated by identically distributed flat Rayleigh fading over one symbol period from a complex Gaussian random variable with zero mean and unit variance and $\textbf{n}_{i}$ is the circularly symmetric white Gaussian noise with zero mean and unit variance at the receiver $R_i$.\\
\indent The first term and the second term in (2) denote the desired signal and the interference signal for $R_{i}$, respectively. Clearly, the desired signals are composed of $K$ received signals and the interference signals are composed of $2K$ received signals.\\
\indent We consider the zero-forcing (ZF) decoder to remove the interference signal at the receiver $R_{i}$. The ZF based decoder can separate  $K \times 3$ X channel into $3K$ P2P channels. Therefore, the sum-rate of the proposed IA scheme is given as
\begin{align*}
R_{sum}=\sum_{\substack{i=1}}^{3}\sum_{j=1}^{K}\text{log}_{2}(1+P_{j}\textbf{R}^{\dagger}_{ij}\textbf{H}_{ij}\textbf{v}_{ij}\textbf{v}^{\dagger}_{ij}\textbf{H}^{\dagger}_{ij}\textbf{R}_{ij}),
\end{align*}
where $\textbf{R}^{\dagger}_{ij}$ is ZF decoder for the desired symbol $s_{ij}$ and $P_{j}$ denotes the transmit power at the transmitter $j$. It is well known that $P_{j}\textbf{R}^{\dagger}_{ij}\textbf{H}_{ij}\textbf{v}_{ij}\textbf{v}^{\dagger}_{ij}\textbf{H}^{\dagger}_{ij}\textbf{R}_{ij}$ is considered as the SNR of the desired signal $s_{ij}$ at the receiver $i$.

\vspace{5pt}

\section{Expanded Beamformer Set Using Latin Square for $K \times 3$ MIMO X Channel}
In this section, we propose an interference alignment scheme for $K \times 3$ MIMO X channel, which can achieve DoF $3K$. At each receiver, $ {M \over 2} (= K) $ DoF can be achieved, which means that the interference signals are aligned in the half of the signal space.

\subsection{Requirements on the Beamforming Vectors}
We propose three requirements on the beamforming vectors in $K \times 3$ MIMO X channel as follows.\\
i)~The $2K$ interference signals at each receiver should be aligned in ${M \over 2}(=K)$ dimensional signal space as
\begin{align}
\mathrm{span}(\textbf{H}_{ij}\textbf{v}_{kj})=\mathrm{span}(\textbf{H}_{im}\textbf{v}_{lm}),\hspace{0.01\textwidth} \text{where}~i,~j,~k,~l,~\text{and}~m~\text{are all distinct}, \hspace{0.01\textwidth} i=1, 2, 3. 
\end{align}
ii)~Any two received interference signals from the same channel should not be aligned along the same dimensional signal space as
\begin{align}
\mathrm{span}(\textbf{H}_{ij}\textbf{v}_{kj})\neq \mathrm{span}(\textbf{H}_{ij}\textbf{v}_{lj}),\hspace{0.01\textwidth} \text{where}~i,~k,~\text{and}~l~\text{are all distinct}, \hspace{0.01\textwidth} i=1, 2, 3. 
\end{align}	
iii)~Each interference pair at each receiver should be aligned as a chain structure, that is,
\begin{align}
&\mathrm{span}(\textbf{H}_{1\alpha_{1}}\textbf{v}_{\delta_{1}})=\mathrm{span}(\textbf{H}_{1\alpha_{2}}\textbf{v}_{\delta_{2}})\hspace{0.01\textwidth}\text{for receiver 1} \nonumber\\
&\mathrm{span}(\textbf{H}_{2\beta_{1}}\textbf{v}_{\delta_{2}})=\mathrm{span}(\textbf{H}_{2\beta_{2}}\textbf{v}_{\delta_{3}})\hspace{0.01\textwidth}\text{for receiver 2}\\
&\mathrm{span}(\textbf{H}_{3\gamma_{1}}\textbf{v}_{\delta_{3}})=\mathrm{span}(\textbf{H}_{3\gamma_{2}}\textbf{v}_{\delta_{1}})\hspace{0.01\textwidth}\text{for receiver 3,}\nonumber
\end{align}
where the requirement iii) can be satisfied in the proposed IA scheme by the Latin square and there are $K$ chain structures.\\
\indent By the requirement i), the $2K$ interference signals received by each receiver are aligned in the $K$ dimensional signal space, which is half of the signal space of each receiver. The requirement ii) means that two beamformers experiencing the same channel are not aligned in the same direction. If the requirement ii) is not satisfied, the corresponding beamformer will be aligned in the same signal space as the desired signal at the other receiver. 
When two interference signals satisfying the requirements i) and ii) are aligned in the same signal space, the available set of beamformers for the IA scheme can be expanded by using the chain structure in (5). In the next section, we will discuss the design of beamformers to satisfy the above three requirements.

\subsection{Design the Beamforming Vectors Using Latin Square}
First, we introduce the Latin square for beamformer design.
The Latin square is a $K \times K$ array filled with $K$ different symbols, each occurring exactly once in each row and exactly once in each column. For example, if the first row is fixed as [A B C], two different $3 \times 3$ Latin squares are given as
\setcounter{MaxMatrixCols}{3}
\begin{equation*}
\begin{bmatrix}
\textbf{A}&\textbf{B}&\textbf{C}\\\textbf{B}&\textbf{C}&\textbf{A}\\\textbf{C}&\textbf{A}&\textbf{B}
\end{bmatrix}
,
\begin{bmatrix}
\textbf{A}&\textbf{B}&\textbf{C}\\\textbf{C}&\textbf{A}&\textbf{B}\\\textbf{B}&\textbf{C}&\textbf{A}
\end{bmatrix}
.
\end{equation*}
\indent Let $\textbf{V}$ be the $K \times 3$ beamforming matrix given as
\begin{equation*}
\textbf{V}=
\begin{bmatrix}
\textbf{v}_{11}&\textbf{v}_{21}&\textbf{v}_{31}\\\textbf{v}_{12}&\textbf{v}_{22}&\textbf{v}_{32}\\\vdots&\vdots&\vdots\\\textbf{v}_{1K}&\textbf{v}_{2K}&\textbf{v}_{3K}
\end{bmatrix}
,
\end{equation*}
where $\textbf{v}_{ij}$ is a $2K \times 1$ beamforming column vector for signal from the $j$-th transmitter to the $i$-th receiver.
The $i$-th column of $\textbf{V}$ represents the beamforming vectors of the signals transmitted to the $i$-th receiver.
That is, in the $i$-th receiver, the remaining two columns except for the $i$-th column represent the beamforming vectors of $2K$ interference signals.
By the requirement i), $2K$ interference signals in each receiver should be aligned in $K$ dimensions so that two interference signals should be aligned in the same dimensional signal space. 
By the requirement ii), the beamformers in the same row should not be aligned in the same signal space because they experience the same channel to the $i$-th receiver.
Also, the beamformers in the same column should not be aligned in the same signal space to have a chain structure of the requirement iii).\\
\indent Considering these conditions, the method to make $K$ pairs of the $2K$ interference signals, where each pair of interference signals are aligned into one signal space is proposed by using three columns in the $K \times K$ Latin square, where one column corresponds to the desired signal and the other two columns correspond to the interference signals. Since the order of aligned interference signals is irrelevant, the order of symbols in the first row of the $K \times K$ Latin square can be fixed. 
Thus, the number of $K \times K$ Latin squares with the first row fixed is $(K-1)!L(K,K)$, where $L(K,K)$ is the value referenced in the online encyclopedia of integer sequences (OEIS) $A000315$, that is, $L(3,3)=1,~L(4,4)=4,~L(5,5)=56,~L(6,6)=9408,~$etc. [15].
The interference alignment pairs of the beamforming vectors are obtained from three columns in the $K \times K$ Latin square with the first row fixed and thus the total number of IA schemes is lower bounded by $(K-1)!L(K,K)$.\\
\indent For simplicity, we design the beamforming vectors for one of the available IA schemes as follows. In each IA scheme with $K$ chain structures, each chain structure in (5) is rewritten as
\begin{align}
\mathrm{span}(\textbf{H}_{3\gamma_{2}}^{-1}\textbf{H}_{3\gamma_{1}}\textbf{H}_{2\beta_{2}}^{-1}&\textbf{H}_{2\beta_{1}}\textbf{H}_{1\alpha_{2}}^{-1}\textbf{H}_{1\alpha_{1}}\textbf{v}_{\delta_{1}})=\mathrm{span}(\textbf{v}_{\delta_{1}})\nonumber\\
&\textbf{v}_{\delta_{2}}=\textbf{H}_{1\alpha_{2}}^{-1}\textbf{H}_{1\alpha_{1}}\textbf{v}_{\delta_{1}}\\
&\textbf{v}_{\delta_{3}}=\textbf{H}_{3\gamma_{1}}^{-1}\textbf{H}_{3\gamma_{2}}\textbf{v}_{\delta_{1}},\nonumber
\end{align}
where $\textbf{v}_{\delta_{1}}$ is obtained from the eigenvectors of $\textbf{H}_{3\gamma_{2}}^{-1}\textbf{H}_{3\gamma_{1}}\textbf{H}_{2\beta_{2}}^{-1}\textbf{H}_{2\beta_{1}}\textbf{H}_{1\alpha_{2}}^{-1}\textbf{H}_{1\alpha_{1}}$ and $\textbf{v}_{\delta_{2}}$ and $\textbf{v}_{\delta_{3}}$ can be obtained as in (6).
Since the size of $\textbf{H}_{3\gamma_{2}}^{-1}\textbf{H}_{3\gamma_{1}}\textbf{H}_{2\beta_{2}}^{-1}\textbf{H}_{2\beta_{1}}\textbf{H}_{1\alpha_{2}}^{-1}\textbf{H}_{1\alpha_{1}}$ is $2K \times 2K$, $\textbf{v}_{\delta_{1}}$ can be selected among $2K$ eigenvectors.
Since there are $K$ chain structures, we have $(2K)^{K}$ eigenvector sets, where each eigenvector set is composed of $K$ eigenvectors. Let $\textbf{B}$ be the beamformer set for each IA scheme and let $\textbf{b}$ be the beamforming vector set, where each element consists of $3K$ beamforming vectors as 
\begin{align} 
\textbf{b}=\{\textbf{v}_{11}, \textbf{v}_{21}, \textbf{v}_{31}, \cdots, \textbf{v}_{1K}, \textbf{v}_{2K}, \textbf{v}_{3K}\}\in\textbf{B}.\nonumber
\end{align}
In each IA scheme, there are $(2K)^{K}$ beamforming vector sets derived from $(2K)^{K}$ eigenvector sets, that is, $|\textbf{B}|=(2K)^{K}$.
Let $\textbf{L}_{K}$ be the expanded beamformer sets by the Latin squares, where each element is beamformer set $\textbf{B}_{i}$.
The proposed IA scheme for $K \times 3$ MIMO X channel has at least $(K-1)!L(K,K)$ available IA schemes and thus we have at least $(K-1)!L(K,K)$ beamformer sets as
\begin{align}
\textbf{L}_{K}=\{\textbf{B}_{i}|i=1,\cdots, (K-1)!L(K,K)\}.
\end{align}
In fact, $(K-1)!L(K,K)$ is large enough for the number of IA schemes and thus we will use it as the number of IA schemes.
Since each beamformer set contains the $(2K)^K$ beamforming vector sets, there are at least $(K-1)!L(K,K)(2K)^K$ expanded beamforming vector sets. 
In order to select the best beamforming vector set, we have to compute performance measure such as symbol error rate (SER) or sum-rate for $(K-1)!L(K,K)(2K)^K$ beamforming vector sets. 
Clearly, it requires tremendous amount of computation.
Thus, we propose how to efficiently select the beamforming vector sets to improve performance of SER and sum-rate for the $K \times 3$ MIMO X channel in Section IV. 

\subsection{Example of the Proposed Scheme: $3 \times 3$ MIMO X Channel}
The proposed IA scheme is applied to the $3 \times 3$ MIMO X channel which satisfies three requirements in (3), (4), and (5).
In the $3 \times 3$ MIMO X channel, the received signals at three receivers are given as
\begin{align}
\textbf{Y}_{1}&=\textbf{H}_{11}\textbf{v}_{11}s_{11}+\textbf{H}_{12}\textbf{v}_{12}s_{12}+\textbf{H}_{13}\textbf{v}_{13}s_{13}\nonumber\\&+\textbf{H}_{11}\textbf{v}_{21}s_{21}+\textbf{H}_{11}\textbf{v}_{31}s_{31}+\textbf{H}_{12}\textbf{v}_{22}s_{22}\nonumber\\&+\textbf{H}_{12}\textbf{v}_{32}s_{32}+\textbf{H}_{13}\textbf{v}_{23}s_{23}+\textbf{H}_{13}\textbf{v}_{33}s_{33}+\textbf{n}_{1}\nonumber\\
\textbf{Y}_{2}&=\textbf{H}_{21}\textbf{v}_{21}s_{21}+\textbf{H}_{22}\textbf{v}_{22}s_{22}+\textbf{H}_{23}\textbf{v}_{23}s_{23}\nonumber\\&+\textbf{H}_{21}\textbf{v}_{11}s_{11}+\textbf{H}_{21}\textbf{v}_{31}s_{31}+\textbf{H}_{22}\textbf{v}_{12}s_{12}\\&+\textbf{H}_{22}\textbf{v}_{32}s_{32}+\textbf{H}_{23}\textbf{v}_{13}s_{13}+\textbf{H}_{23}\textbf{v}_{33}s_{33}+\textbf{n}_{2}\nonumber\\
\textbf{Y}_{3}&=\textbf{H}_{31}\textbf{v}_{31}s_{31}+\textbf{H}_{32}\textbf{v}_{32}s_{32}+\textbf{H}_{33}\textbf{v}_{33}s_{33}\nonumber\\&+\textbf{H}_{31}\textbf{v}_{11}s_{11}+\textbf{H}_{31}\textbf{v}_{21}s_{21}+\textbf{H}_{32}\textbf{v}_{12}s_{12}\nonumber\\&+\textbf{H}_{32}\textbf{v}_{22}s_{22}+\textbf{H}_{33}\textbf{v}_{13}s_{13}+\textbf{H}_{33}\textbf{v}_{23}s_{23}+\textbf{n}_{3}.\nonumber
\end{align}
Six interference signals are received at each receiver and each pair of interference signals are aligned in the same dimensional signal space to satisfy the requirement i). The choice for two interference signals aligning into one dimensional signal space can be determined by the Latin square, where there are two distinct $3 \times 3$ Latin squares. In other words, interference alignment pair can be determined by the $3 \times 3$ Latin squares as
\begin{equation}
\begin{bmatrix}
\textbf{v}_{11}&\textbf{v}_{21}&\textbf{v}_{31}\\\textbf{v}_{12}&\textbf{v}_{22}&\textbf{v}_{32}\\\textbf{v}_{13}&\textbf{v}_{23}&\textbf{v}_{33}
\end{bmatrix}
\longleftarrow
\begin{bmatrix}
\textbf{A}&\textbf{B}&\textbf{C}\\\textbf{B}&\textbf{C}&\textbf{A}\\\textbf{C}&\textbf{A}&\textbf{B}
\end{bmatrix}
\text{or}
\begin{bmatrix}
\textbf{A}&\textbf{B}&\textbf{C}\\\textbf{C}&\textbf{A}&\textbf{B}\\\textbf{B}&\textbf{C}&\textbf{A}
\end{bmatrix}
.
\end{equation}
Each transmitter beamforms to align two interference signals into one dimensional signal space at each receiver.
For example, the signals $s_{11}$, $s_{12}$, and $s_{13}$ are desired signals at the receiver 1 and the other six signals are interference signals in (8).
Thus, the first column of the beamforming matrix in (9) is the beamforming vectors for the desired signals of the receiver 1.
Each pair of interference signals corresponding to each pair of the same symbols in the second and the third columns in the Latin squares in (9) are aligned in the one dimensional interference signal space, that is, for the first Latin square, each pair of $(\textbf{v}_{23}, \textbf{v}_{32})$ from $\textbf{A}$, $(\textbf{v}_{21}, \textbf{v}_{33})$ from $\textbf{B}$, and $(\textbf{v}_{22}, \textbf{v}_{31})$ from $\textbf{C}$ is aligned in the one dimensional interference signal space. That is, the interference signal pairs of three receivers can be obtained from 
\begin{equation*}
\begin{bmatrix}
\textbf{v}_{21}&\textbf{v}_{31}\\\textbf{v}_{22}&\textbf{v}_{32}\\\textbf{v}_{23}&\textbf{v}_{33}
\end{bmatrix}
\longleftarrow
\begin{bmatrix}
\textbf{B}&\textbf{C}\\\textbf{C}&\textbf{A}\\\textbf{A}&\textbf{B}
\end{bmatrix}
\text{for receiver 1}
\end{equation*}
\begin{equation*}
\begin{bmatrix}
\textbf{v}_{11}&\textbf{v}_{31}\\\textbf{v}_{12}&\textbf{v}_{32}\\\textbf{v}_{13}&\textbf{v}_{33}
\end{bmatrix}
\longleftarrow
\begin{bmatrix}
\textbf{A}&\textbf{C}\\\textbf{B}&\textbf{A}\\\textbf{C}&\textbf{B}
\end{bmatrix}
\text{for receiver 2}
\end{equation*}
\begin{equation*}
\begin{bmatrix}
\textbf{v}_{11}&\textbf{v}_{21}\\\textbf{v}_{12}&\textbf{v}_{22}\\\textbf{v}_{13}&\textbf{v}_{23}
\end{bmatrix}
\longleftarrow
\begin{bmatrix}
\textbf{A}&\textbf{B}\\\textbf{B}&\textbf{C}\\\textbf{C}&\textbf{A}
\end{bmatrix}
\text{for receiver 3}
\end{equation*}
and thus the three pairs of interference signals at each receiver are given as
\begin{align*}
\text{Receiver 1;}&~(\textbf{v}_{23},\textbf{v}_{32}),~(\textbf{v}_{21},\textbf{v}_{33}),~(\textbf{v}_{22},\textbf{v}_{31})\\
\text{Receiver 2;}&~(\textbf{v}_{11},\textbf{v}_{32}),~(\textbf{v}_{12},\textbf{v}_{33}),~(\textbf{v}_{13},\textbf{v}_{31})\\
\text{Receiver 3;}&~(\textbf{v}_{11},\textbf{v}_{23}),~(\textbf{v}_{12},\textbf{v}_{21}),~(\textbf{v}_{13},\textbf{v}_{22}).
\end{align*}

\indent By aligning the interference signals in the form of the first Latin square in (9), the following IA condition is obtained from (8) as
\begin{align}
\mathrm{span}(\textbf{H}_{31}\textbf{v}_{11})=\mathrm{span}(\textbf{H}_{33}\textbf{v}_{23})\nonumber\\
\mathrm{span}(\textbf{H}_{21}\textbf{v}_{11})=\mathrm{span}(\textbf{H}_{22}\textbf{v}_{32})\nonumber\\
\mathrm{span}(\textbf{H}_{13}\textbf{v}_{23})=\mathrm{span}(\textbf{H}_{12}\textbf{v}_{32})\nonumber
\end{align}
\begin{align}
\mathrm{span}(\textbf{H}_{31}\textbf{v}_{21})=\mathrm{span}(\textbf{H}_{32}\textbf{v}_{12})\nonumber\\
\mathrm{span}(\textbf{H}_{11}\textbf{v}_{21})=\mathrm{span}(\textbf{H}_{13}\textbf{v}_{33})\\
\mathrm{span}(\textbf{H}_{22}\textbf{v}_{12})=\mathrm{span}(\textbf{H}_{23}\textbf{v}_{33})\nonumber
\end{align}
\begin{align}
\mathrm{span}(\textbf{H}_{21}\textbf{v}_{31})=\mathrm{span}(\textbf{H}_{23}\textbf{v}_{13})\nonumber\\
\mathrm{span}(\textbf{H}_{11}\textbf{v}_{31})=\mathrm{span}(\textbf{H}_{12}\textbf{v}_{22})\nonumber\\
\mathrm{span}(\textbf{H}_{33}\textbf{v}_{13})=\mathrm{span}(\textbf{H}_{32}\textbf{v}_{22}).\nonumber
\end{align}
Using (10), the beamforming vectors for the IA scheme are obtained from
\begin{align}
\mathrm{span}(\textbf{H}_{21}^{-1}\textbf{H}_{22}\textbf{H}_{12}^{-1}&\textbf{H}_{13}\textbf{H}_{33}^{-1}\textbf{H}_{31}\textbf{v}_{11})=\mathrm{span}(\textbf{v}_{11})\nonumber\\
&\textbf{v}_{23}=\textbf{H}_{33}^{-1}\textbf{H}_{31}\textbf{v}_{11}\\
&\textbf{v}_{32}=\textbf{H}_{22}^{-1}\textbf{H}_{21}\textbf{v}_{11}\nonumber
\end{align}
\begin{align}
\mathrm{span}(\textbf{H}_{11}^{-1}\textbf{H}_{13}\textbf{H}_{23}^{-1}&\textbf{H}_{22}\textbf{H}_{32}^{-1}\textbf{H}_{31}\textbf{v}_{21})=\mathrm{span}(\textbf{v}_{21})\nonumber\\
&\textbf{v}_{12}=\textbf{H}_{32}^{-1}\textbf{H}_{31}\textbf{v}_{21}\\
&\textbf{v}_{33}=\textbf{H}_{13}^{-1}\textbf{H}_{11}\textbf{v}_{21}\nonumber
\end{align}
\begin{align}
\mathrm{span}(\textbf{H}_{11}^{-1}\textbf{H}_{12}\textbf{H}_{32}^{-1}&\textbf{H}_{33}\textbf{H}_{23}^{-1}\textbf{H}_{21}\textbf{v}_{31})=\mathrm{span}(\textbf{v}_{31})\nonumber\\
&\textbf{v}_{13}=\textbf{H}_{23}^{-1}\textbf{H}_{21}\textbf{v}_{31}\\
&\textbf{v}_{22}=\textbf{H}_{12}^{-1}\textbf{H}_{11}\textbf{v}_{31}.\nonumber
\end{align}
Clearly, $\textbf{v}_{11}$ is obtained from the eigenvectors of $\textbf{H}_{21}^{-1}\textbf{H}_{22}\textbf{H}_{12}^{-1}\textbf{H}_{13}\textbf{H}_{33}^{-1}\textbf{H}_{31}(=\textbf{E}_{1})$ and $\textbf{v}_{23}$ and $\textbf{v}_{32}$ are also obtained from (11). Also, $\textbf{v}_{21}$ and $\textbf{v}_{31}$ are obtained from the eigenvectors of $\textbf{H}_{11}^{-1}\textbf{H}_{13}\textbf{H}_{23}^{-1}\textbf{H}_{22}\textbf{H}_{32}^{-1}\textbf{H}_{31}(=\textbf{E}_{2})$ and $\textbf{H}_{11}^{-1}\textbf{H}_{12}\textbf{H}_{32}^{-1}\textbf{H}_{33}\textbf{H}_{23}^{-1}\textbf{H}_{21}(=\textbf{E}_{3})$, respectively. Therefore, the nine beamforming vectors can be obtained from (11), (12), and  (13). Since the size of $\textbf{E}_{i}$ is $6 \times 6$, the number of eigenvectors that can be selected for each of $\textbf{v}_{11}$, $\textbf{v}_{21}$, and $\textbf{v}_{31}$ is 6. There are $6^{3}$ beamforming vector sets derived from $6^{3}$ eigenvector sets. Thus, the selectable number of beamforming vector sets for the IA scheme is $6^3$.\\
\indent Also, by aligning the interference signals in the form of the second Latin square matrix in (9), the following IA condition is obtained as three chain structures
\begin{align}
\mathrm{span}(\textbf{H}_{31}\textbf{v}_{11})=\mathrm{span}(\textbf{H}_{32}\textbf{v}_{22})\nonumber\\
\mathrm{span}(\textbf{H}_{21}\textbf{v}_{11})=\mathrm{span}(\textbf{H}_{23}\textbf{v}_{33})\nonumber\\
\mathrm{span}(\textbf{H}_{12}\textbf{v}_{22})=\mathrm{span}(\textbf{H}_{13}\textbf{v}_{33})\nonumber
\end{align}
\begin{align}
\mathrm{span}(\textbf{H}_{11}\textbf{v}_{21})=\mathrm{span}(\textbf{H}_{12}\textbf{v}_{32})\nonumber\\
\mathrm{span}(\textbf{H}_{31}\textbf{v}_{21})=\mathrm{span}(\textbf{H}_{33}\textbf{v}_{13})\\
\mathrm{span}(\textbf{H}_{22}\textbf{v}_{32})=\mathrm{span}(\textbf{H}_{23}\textbf{v}_{13})\nonumber
\end{align}
\begin{align}
\mathrm{span}(\textbf{H}_{11}\textbf{v}_{31})=\mathrm{span}(\textbf{H}_{13}\textbf{v}_{23})\nonumber\\
\mathrm{span}(\textbf{H}_{21}\textbf{v}_{31})=\mathrm{span}(\textbf{H}_{22}\textbf{v}_{12})\nonumber\\
\mathrm{span}(\textbf{H}_{32}\textbf{v}_{12})=\mathrm{span}(\textbf{H}_{33}\textbf{v}_{23})\nonumber
\end{align}
and we have
\begin{align}
\mathrm{span}(\textbf{H}_{21}^{-1}\textbf{H}_{23}\textbf{H}_{13}^{-1}&\textbf{H}_{12}\textbf{H}_{32}^{-1}\textbf{H}_{31}\textbf{v}_{11})=\mathrm{span}(\textbf{v}_{11})\nonumber\\
&\textbf{v}_{22}=\textbf{H}_{32}^{-1}\textbf{H}_{31}\textbf{v}_{11}\\
&\textbf{v}_{33}=\textbf{H}_{23}^{-1}\textbf{H}_{21}\textbf{v}_{11}\nonumber
\end{align}
\begin{align}
\mathrm{span}(\textbf{H}_{31}^{-1}\textbf{H}_{33}\textbf{H}_{23}^{-1}&\textbf{H}_{22}\textbf{H}_{12}^{-1}\textbf{H}_{11}\textbf{v}_{21})=\mathrm{span}(\textbf{v}_{21})\nonumber\\
&\textbf{v}_{13}=\textbf{H}_{33}^{-1}\textbf{H}_{31}\textbf{v}_{21}\\
&\textbf{v}_{32}=\textbf{H}_{12}^{-1}\textbf{H}_{11}\textbf{v}_{21}\nonumber
\end{align}
\begin{align}
\mathrm{span}(\textbf{H}_{11}^{-1}\textbf{H}_{13}\textbf{H}_{33}^{-1}&\textbf{H}_{32}\textbf{H}_{22}^{-1}\textbf{H}_{21}\textbf{v}_{31})=\mathrm{span}(\textbf{v}_{31})\nonumber\\
&\textbf{v}_{12}=\textbf{H}_{22}^{-1}\textbf{H}_{21}\textbf{v}_{31}\\
&\textbf{v}_{23}=\textbf{H}_{13}^{-1}\textbf{H}_{11}\textbf{v}_{31}.\nonumber
\end{align}
Similarly, the nine beamforming vectors can also be obtained from (15), (16), and (17) and the selectable number of beamforming vector sets for the IA scheme is also $6^3$. In the $3 \times 3$ MIMO X channel, there are $2$ available IA schemes, where each IA scheme has $6^3$ beamforming vector sets. Thus, there are $2 \times 6^3$ expanded beamforming vector sets by two Latin squares.

\subsection{Example of the Proposed Scheme: $4 \times 3$ MIMO X Channel}
For $4 \times 3$ MIMO X channel, one of the available IA schemes is considered.
Eight interference signals are received at each receiver and each pair of interference signals are aligned in the same dimensional signal space to satisfy the requirement i). The choice for two interference signals aligning into one dimensional signal space can be determined by the $4 \times 4$ Latin square, where there are at least $3!$ Latin squares. Interference alignment pair can be determined by any three columns of the $4 \times 4$ Latin squares given as

\begin{equation}
\begin{bmatrix}
\textbf{v}_{11}&\textbf{v}_{21}&\textbf{v}_{31}\\\textbf{v}_{12}&\textbf{v}_{22}&\textbf{v}_{32}\\\textbf{v}_{13}&\textbf{v}_{23}&\textbf{v}_{33}\\\textbf{v}_{14}&\textbf{v}_{24}&\textbf{v}_{34}
\end{bmatrix}
\longleftarrow
\begin{bmatrix}
\textbf{A}&\textbf{B}&\textbf{C}\\\textbf{B}&\textbf{C}&\textbf{D}\\\textbf{C}&\textbf{D}&\textbf{A}\\\textbf{D}&\textbf{A}&\textbf{B}
\end{bmatrix}
,
\end{equation}
where one column corresponds to the desired signal and the other two columns correspond to the interference signals. 
Each transmitter beamforms to align two interference signals into one dimensional signal space at each receiver.
Thus, the first column of the beamforming matrix is the beamforming vectors for the desired signals of the receiver 1.
Each pair of interference signals corresponding to each pair of the same symbols in the second and the third columns are aligned in the one dimensional interference signal space, that is, for the Latin square matrix in (18), each pair of $(\textbf{v}_{24}, \textbf{v}_{33})$ from $\textbf{A}$, $(\textbf{v}_{21}, \textbf{v}_{34})$ from $\textbf{B}$, $(\textbf{v}_{22}, \textbf{v}_{31})$ from $\textbf{C}$, and $(\textbf{v}_{23}, \textbf{v}_{32})$ from $\textbf{D}$ is aligned in the one dimensional interference signal space at the receiver 1.\\
\indent By aligning the interference signals in the form of the Latin square matrix in (18), the following IA condition is obtained as four chain structures
\begin{align}
\mathrm{span}(\textbf{H}_{14}\textbf{v}_{24})=\mathrm{span}(\textbf{H}_{13}\textbf{v}_{33})\nonumber\\
\mathrm{span}(\textbf{H}_{21}\textbf{v}_{11})=\mathrm{span}(\textbf{H}_{23}\textbf{v}_{33})\nonumber\\
\mathrm{span}(\textbf{H}_{31}\textbf{v}_{11})=\mathrm{span}(\textbf{H}_{34}\textbf{v}_{24})\nonumber
\end{align}
\begin{align}
\mathrm{span}(\textbf{H}_{11}\textbf{v}_{21})=\mathrm{span}(\textbf{H}_{14}\textbf{v}_{34})\nonumber\\
\mathrm{span}(\textbf{H}_{22}\textbf{v}_{12})=\mathrm{span}(\textbf{H}_{24}\textbf{v}_{34})\\
\mathrm{span}(\textbf{H}_{32}\textbf{v}_{12})=\mathrm{span}(\textbf{H}_{31}\textbf{v}_{21})\nonumber
\end{align}
\begin{align}
\mathrm{span}(\textbf{H}_{12}\textbf{v}_{22})=\mathrm{span}(\textbf{H}_{11}\textbf{v}_{31})\nonumber\\
\mathrm{span}(\textbf{H}_{23}\textbf{v}_{13})=\mathrm{span}(\textbf{H}_{21}\textbf{v}_{31})\nonumber\\
\mathrm{span}(\textbf{H}_{33}\textbf{v}_{13})=\mathrm{span}(\textbf{H}_{32}\textbf{v}_{22})\nonumber
\end{align}
\begin{align}
&\mathrm{span}(\textbf{H}_{13}\textbf{v}_{23})=\mathrm{span}(\textbf{H}_{12}\textbf{v}_{32})\nonumber\\
&\mathrm{span}(\textbf{H}_{24}\textbf{v}_{14})=\mathrm{span}(\textbf{H}_{22}\textbf{v}_{32})\nonumber\\
&\mathrm{span}(\textbf{H}_{34}\textbf{v}_{14})=\mathrm{span}(\textbf{H}_{33}\textbf{v}_{23}).\nonumber
\end{align}

Using (19), the beamforming vectors for the IA scheme are obtained from
\begin{align}
\mathrm{span}(\textbf{H}_{31}^{-1}\textbf{H}_{34}\textbf{H}_{14}^{-1}&\textbf{H}_{13}\textbf{H}_{23}^{-1}\textbf{H}_{21}\textbf{v}_{11})=\mathrm{span}(\textbf{v}_{11})\nonumber\\
&\textbf{v}_{33}=\textbf{H}_{23}^{-1}\textbf{H}_{21}\textbf{v}_{11}\\
&\textbf{v}_{24}=\textbf{H}_{34}^{-1}\textbf{H}_{31}\textbf{v}_{11}\nonumber
\end{align}
\begin{align}
\mathrm{span}(\textbf{H}_{31}^{-1}\textbf{H}_{32}\textbf{H}_{22}^{-1}&\textbf{H}_{24}\textbf{H}_{14}^{-1}\textbf{H}_{11}\textbf{v}_{21})=\mathrm{span}(\textbf{v}_{21})\nonumber\\
&\textbf{v}_{34}=\textbf{H}_{14}^{-1}\textbf{H}_{11}\textbf{v}_{21}\\
&\textbf{v}_{12}=\textbf{H}_{32}^{-1}\textbf{H}_{31}\textbf{v}_{21}\nonumber
\end{align}
\begin{align}
\mathrm{span}(\textbf{H}_{21}^{-1}\textbf{H}_{23}\textbf{H}_{33}^{-1}&\textbf{H}_{32}\textbf{H}_{12}^{-1}\textbf{H}_{11}\textbf{v}_{31})=\mathrm{span}(\textbf{v}_{31})\nonumber\\
&\textbf{v}_{22}=\textbf{H}_{12}^{-1}\textbf{H}_{11}\textbf{v}_{31}\\
&\textbf{v}_{13}=\textbf{H}_{23}^{-1}\textbf{H}_{21}\textbf{v}_{31}\nonumber
\end{align}
\begin{align}
\mathrm{span}(\textbf{H}_{34}^{-1}\textbf{H}_{33}\textbf{H}_{13}^{-1}&\textbf{H}_{12}\textbf{H}_{22}^{-1}\textbf{H}_{24}\textbf{v}_{14})=\mathrm{span}(\textbf{v}_{14})\nonumber\\
&\textbf{v}_{32}=\textbf{H}_{22}^{-1}\textbf{H}_{24}\textbf{v}_{14}\\
&\textbf{v}_{23}=\textbf{H}_{33}^{-1}\textbf{H}_{34}\textbf{v}_{14}.\nonumber
\end{align}
Here, $\textbf{v}_{11}$ is obtained from the eigenvectors of $\textbf{H}_{31}^{-1}\textbf{H}_{34}\textbf{H}_{14}^{-1}\textbf{H}_{13}\textbf{H}_{23}^{-1}\textbf{H}_{21}(=\textbf{E}_{1})$ and $\textbf{v}_{21}$, $\textbf{v}_{31}$, and $\textbf{v}_{14}$ are obtained from the eigenvectors of $\textbf{H}_{31}^{-1}\textbf{H}_{32}\textbf{H}_{22}^{-1}\textbf{H}_{24}\textbf{H}_{14}^{-1}\textbf{H}_{11}(=\textbf{E}_{2})$,\\ $\textbf{H}_{21}^{-1}\textbf{H}_{23}\textbf{H}_{33}^{-1}\textbf{H}_{32}\textbf{H}_{12}^{-1}\textbf{H}_{11}(=\textbf{E}_{3})$, and $\textbf{H}_{34}^{-1}\textbf{H}_{33}\textbf{H}_{13}^{-1}\textbf{H}_{12}\textbf{H}_{22}^{-1}\textbf{H}_{24}(=\textbf{E}_{4})$, respectively. Therefore, the 12 beamforming vectors can be obtained from (20), (21), (22), and (23). Since the size of $\textbf{E}_{i}$ is $8 \times 8$, the number of eigenvectors that can be selected for each of $\textbf{v}_{11}$, $\textbf{v}_{21}$, $\textbf{v}_{31}$, and $\textbf{v}_{14}$ is 8. There are $8^{4}$ beamforming vector sets derived from $8^{4}$ eigenvector sets. Thus, the selectable number of beamforming vector sets for the IA scheme is $8^4$. In the $4 \times 3$ MIMO X channel, there are at least $3!$ IA schemes, where each IA scheme has $8^4$ beamforming vector sets. Therefore, there are $3!8^4$ expanded beamforming vector sets by the Latin squares.

\vspace{5pt}

\section{Efficient Beamformer Selection Methods}
We can select the beamforming vector set $\textbf{b}^*\in\textbf{B}$ by using the following methods:
\begin{itemize}
\item MinMax SNR based beamformer selection method;
\begin{align}
\textbf{b}^*=\operatorname{arg}\underset{\textbf{b}\in\textbf{B}}{\operatorname{max}}\underset{i}{\operatorname{min}}|(\textbf{R}^{\dagger}\textbf{H}\textbf{v})_{i}|,\hspace{0.03\textwidth} i=1, 2, \cdots, 3K,  
\end{align}
where $\textbf{R}^{\dagger}$, $\textbf{H}$, and $\textbf{v}$ denote the zero-forcing matirix, the channel matrix, and the beamforming vector for the desired signal $i$, respectively.\\
\item Sum-rate maximization based beamformer selection method;
\begin{align}
\textbf{b}^*=\operatorname{arg}\underset{\textbf{b}\in\textbf{B}}{\operatorname{max}}(R_{sum}),  
\end{align}
where $R_{sum}$ denotes the sum-rate at the given system.\\
\item Condition number (CN) based beamformer selection method;
\begin{align}
\textbf{b}^*=\operatorname{arg}\underset{\textbf{b}\in\textbf{B}}{\operatorname{min}}(F(\kappa(\textbf{b}))),  
\end{align}
where $\kappa(\textbf{b})$ is the CN for the IA scheme with \textbf{b} and $F(\cdot)$ is a function mapping from CN to MinMax SNR or sum-rate.
\end{itemize}

\subsection{Utilization of Condition Number}
Instead of measuring the SNR or sum-rate with high computational complexity, we introduce other measurements for the beamformer selection deriving suboptimal solutions with less computational complexity. That is, we use the CN as a measurement to estimate SNR or sum-rate for the $K \times 3$ MIMO X channel. The CN is defined as a ratio of the maximum singular value to the minimum singular value, where the singular value is a weight that indicates how much the range space in the matrix is biased in the given direction. That is, the CN is the ratio of the singular value in the most biased direction to that in the least biased direction of the matrix.
When the CN reaches the minimum value of 1, it becomes an orthogonal matrix. Therefore, the closer the CN is to 1 (the smaller), the closer to the orthogonality of the matrix. In general, it is known that the orthogonalization of the received signals can achieve the improvement of performance in the interference channel.\\ 
\indent For each receiver, the signal space matrix is defined as 
\begin{align}
\textbf{A}_{i}=[\textbf{D}_{1}, \textbf{D}_{2}, \cdots, \textbf{D}_{K}, \textbf{I}_{1}, \textbf{I}_{2}, \cdots, \textbf{I}_{K}], \hspace{0.03\textwidth} i=1, 2, 3,
\end{align}
where $\textbf{D}_{i}$ and $\textbf{I}_{i}$ are $2K \times 1$ desired signal vectors and aligned interference signal vectors, respectively. Each receiver decodes the desired signal by zero-forcing. However, the desired signal is attenuated by zero-forcing, which is measured by the CN of the signal space matrix in (27). In order to consider the orthogonality relation rather than the magnitude of the signal space matrix, we use a normalized signal space matrix by normalizing each signal for an orthogonality measure as
\begin{align}
\bar{\textbf{A}}_{i}=[{\textbf{D}_{1}\over |\textbf{D}_{1}|}, {\textbf{D}_{2}\over |\textbf{D}_{2}|}, \cdots, {\textbf{D}_{K}\over |\textbf{D}_{K}|}, {\textbf{I}_{1}\over |\textbf{I}_{1}|}, {\textbf{I}_{2}\over |\textbf{I}_{2}|}, \cdots, {\textbf{I}_{K}\over |\textbf{I}_{K}|}], \hspace{0.03\textwidth} i=1, 2, 3.
\end{align}
Then, the CN is obtained for each normalized signal space matrix of each receiver and used as a measure of the performance evaluation, whose computational complexity can be reduced compared to the SNR or sum-rate computation.

\subsection{Correlation between Condition Number and Minimum SNR}
For the case of $3 \times 3$ MIMO X channel, there are $|\textbf{B}|=6^{3}$ beamforming vector sets for an IA scheme.
For each beamforming vector set, the minimum SNR and the estimated minimum SNR from CN are computed as follows:\begin{itemize}
\item Find the minimum SNR.
\begin{itemize}
\item For each of the $6^{3}$ beamforming vector sets, sort $\underset{i}{\operatorname{min}}|(\textbf{R}^{\dagger}\textbf{H}\textbf{v})_{i}|^2,\hspace{0.01\textwidth} i=1, 2, \cdots, 9$, in the descending order.
\end{itemize}
\item Estimate the minimum SNR from CN.
\begin{itemize}
\item For each $\textbf{b}_{j}\in\textbf{B}$, compute CN by $\underset{i\in(1,2,3)}{\operatorname{max}}(\kappa_{i}(\textbf{b}_{j})),\hspace{0.01\textwidth}j=1, 2, \cdots, 6^{3}$ of the normalized signal space matrix obtained from $6^{3}$ beamforming vector sets, where $\kappa_{i}(\textbf{b}_{j})$ is the CN of the $i$-th receiver with the beamforming vector set $\textbf{b}_{j}$ in \textbf{B}.
\item Let $\bar{\textbf{b}}_{j}$ be the sorted beamforming vector in the ascending order of $\underset{i}{\operatorname{max}}(\kappa_{i}(\textbf{b}_{j}))$. 
\item For each of $\bar{\textbf{b}}_{j}$, obtain $\underset{i}{\operatorname{min}}|(\textbf{R}^{\dagger}\textbf{H}\textbf{v})_{i}|^2,\hspace{0.01\textwidth} i=1, 2, \cdots, 9$.
\end{itemize}
\item Compare the minimum SNR and the estimated minimum SNR from CN.
\end{itemize}
\begin{figure}[t!]
		\centering
    \includegraphics[width=0.6\textwidth]{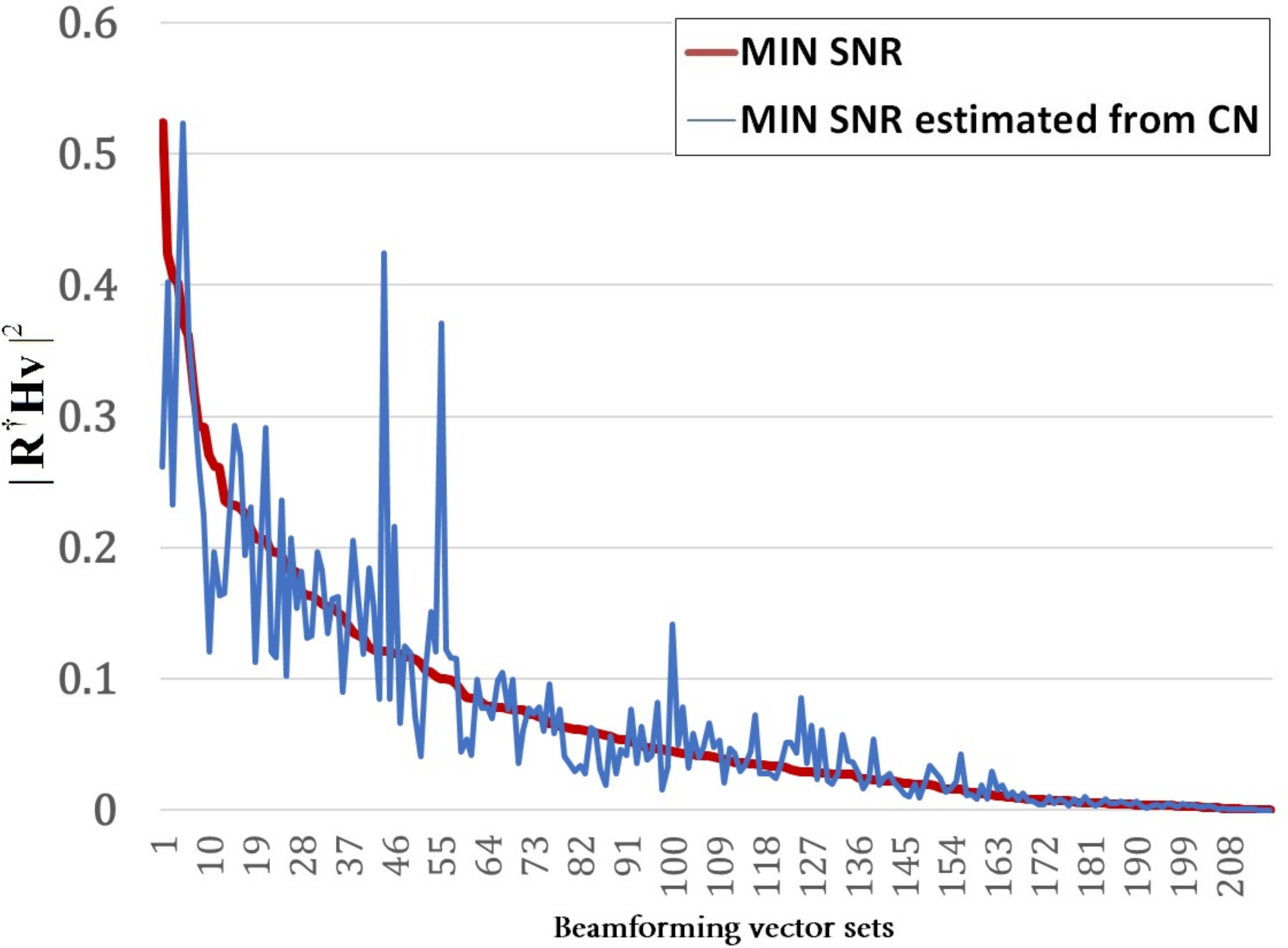}
		\caption{Comparison of the directly obtained minimum SNR and the estimated minimum SNR from CN in the $3 \times 3$ X channel.}
\end{figure}
\indent Some reasons for the fluctuation in Fig. 2 are as follows. First, CN can be considered as a measurement of orthogonality among received signals. Second, we need only the orthogonality between the desired signal and the interference signal. But CN also computes a measure of orthogonality among the interference signals. By considering the beamforming vector sets within a certain range of the minimum SNR estimated from CN, it is possible to compromise the fluctuation in estimating the minimum SNR by CN as follows: 
\begin{itemize}
\item Choose a set of $\bar{\textbf{b}}_{j}$'s with the smallest $u$ CN's and compute the minimum SNR of each $\bar{\textbf{b}}_{j}$ among them. 
\item Select the $\bar{\textbf{b}}^{*}$ that gives the maximum SNR among them.
\end{itemize}
\indent Fig. 3 compares the MinMax SNR distribution for the sets of $\bar{\textbf{b}}_{j}$ with the smallest CN's, denoted by CN$u$ with set size $u$, $u=1,~3,~10,~20$.
For CN10, the shape of the SNR distribution is similar to that of the optimal case and it is nearly optimal for CN20. Thus, MinMax SNR by CN20 is almost the same as the Optimal MinMax SNR.
\begin{figure}[t!]
		\centering
    \includegraphics[width=0.6\textwidth]{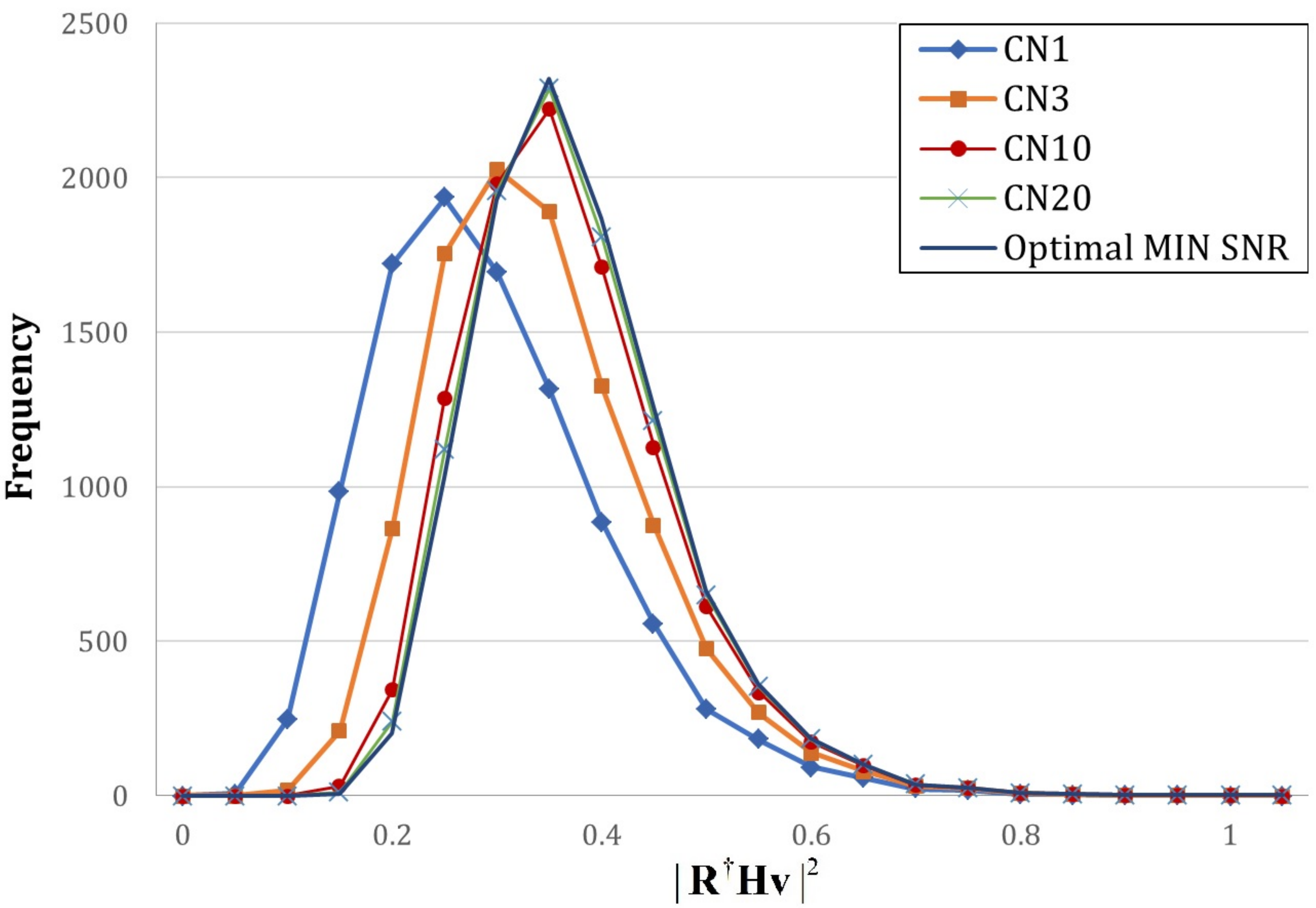}
		\caption{Comparison of the MinMax SNR distribution for various CN$u$ in the $3 \times 3$ X channel.}
\end{figure}
\subsection{Interference Orthogonalized Condition Number}
\begin{figure}[t!]
		\centering
    \includegraphics[width=0.6\textwidth]{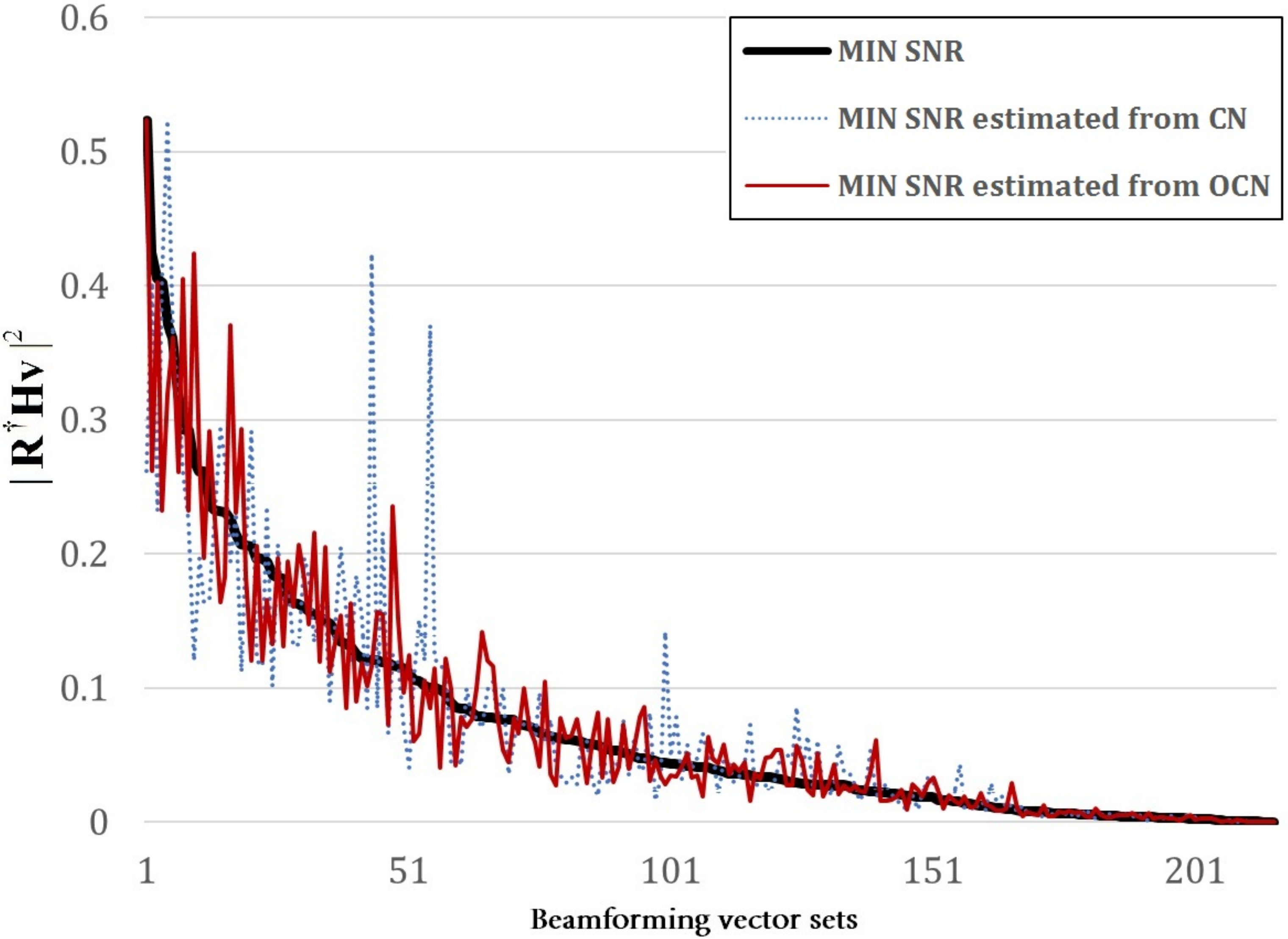}
		\caption{Comparison of the directly obtained minimum SNR and the estimated minimum SNRs from CN and OCN in the $3 \times 3$ X channel.}
\label{beam3}
\end{figure}

With a large number of users, the number of beamforming vector sets increases exponentially and thus the set size CN$u$ should be large. Thus we propose a modified CN that improves the method to obtain CN from signal space. In order to avoid computation of the orthogonality between the interference signals, a new signal space matrix can be made, which considers only the orthogonality between the desired and interference signals. Thus, the signal space matrix can be constructed by pre-orthogonalizing the interference signal space $[{\textbf{I}_{1}\over |\textbf{I}_{1}|}, {\textbf{I}_{2}\over |\textbf{I}_{2}|}, \cdots, {\textbf{I}_{K}\over |\textbf{I}_{K}|}]$ in the signal space matrix in (28) using Gram-Schmidt orthogonalization. Then, its condition number is called orthogonalized condition number (OCN), which is useful for the large $K$.\\ 
\indent In Fig. \ref{beam3}, compared to the fluctuation of the estimated minimum SNR from CN, the fluctuation of the estimated minimum SNR from the OCN is reduced.

\subsection{Correlation between Condition Number and Sum-Rate}
\indent As in the previous case, the procedure to obtain the maximum sum-rate among all beamforming vector sets can be replaced by the CN based beamformer selection with less computational complexity.\\ 
\indent For the case of $3 \times 3$ MIMO X channel, there are $|\textbf{B}|=6^{3}$ beamforming vector sets for an IA scheme.
For each beamforming vector set, the maximum sum-rate and the estimated maximum sum-rate from CN are computed as follows:
\begin{itemize}
\item Find the sum-rate.
\begin{itemize}
\item For the $6^{3}$ beamforming vector sets, sort the sum-rates for all beamforming vector sets in the descending order.
\end{itemize}
\item Estimate the sum-rate from CN.
\begin{itemize}
\item For each $\textbf{b}_{j}\in\textbf{B}$, compute CN by $\sum_{i=1}^{3}(\kappa_{i}(\textbf{b}_{j}))$, $\hspace{0.01\textwidth}j=1, 2, \cdots, 6^{3}$ of the normalized signal space matrix obtained from $6^{3}$ beamforming vector sets, where $\kappa_{i}(\textbf{b}_{j})$ is the CN of the $i$-th receiver with the beamforming vector set $\textbf{b}_{j}$ in \textbf{B}.
\item Let $\bar{\textbf{b}}_{j}$ be the sorted beamforming vector in the ascending order of $\sum_{i=1}^{3}(\kappa_{i}(\textbf{b}_{j}))$.
\item For each of $\bar{\textbf{b}}_{j}$, obtain the sum-rate.
\end{itemize}
\item Compare the sum-rate and the estimated sum-rate from CN.
\end{itemize}
Comparing the directly obtained sum-rate with the estimated sum-rate from CN, there are some fluctuations for estimating the maximum sum-rate from CN as in Fig. 5. By considering the beamforming vector sets within a certain range of the sum-rate estimated from CN, it is possible to compromise the fluctuation in estimating the maximum sum-rate by CN as follows: 
\begin{itemize}
\item Choose a set of $\bar{\textbf{b}}_{j}$ with the smallest CN's and compute the sum-rate of each $\bar{\textbf{b}}_{j}$ among them. 
\item Select the $\bar{\textbf{b}}^{*}$ that gives the maximum sum-rate among them.
\end{itemize}
\begin{figure}[t!]
\centering
\includegraphics[width=0.5\textwidth]{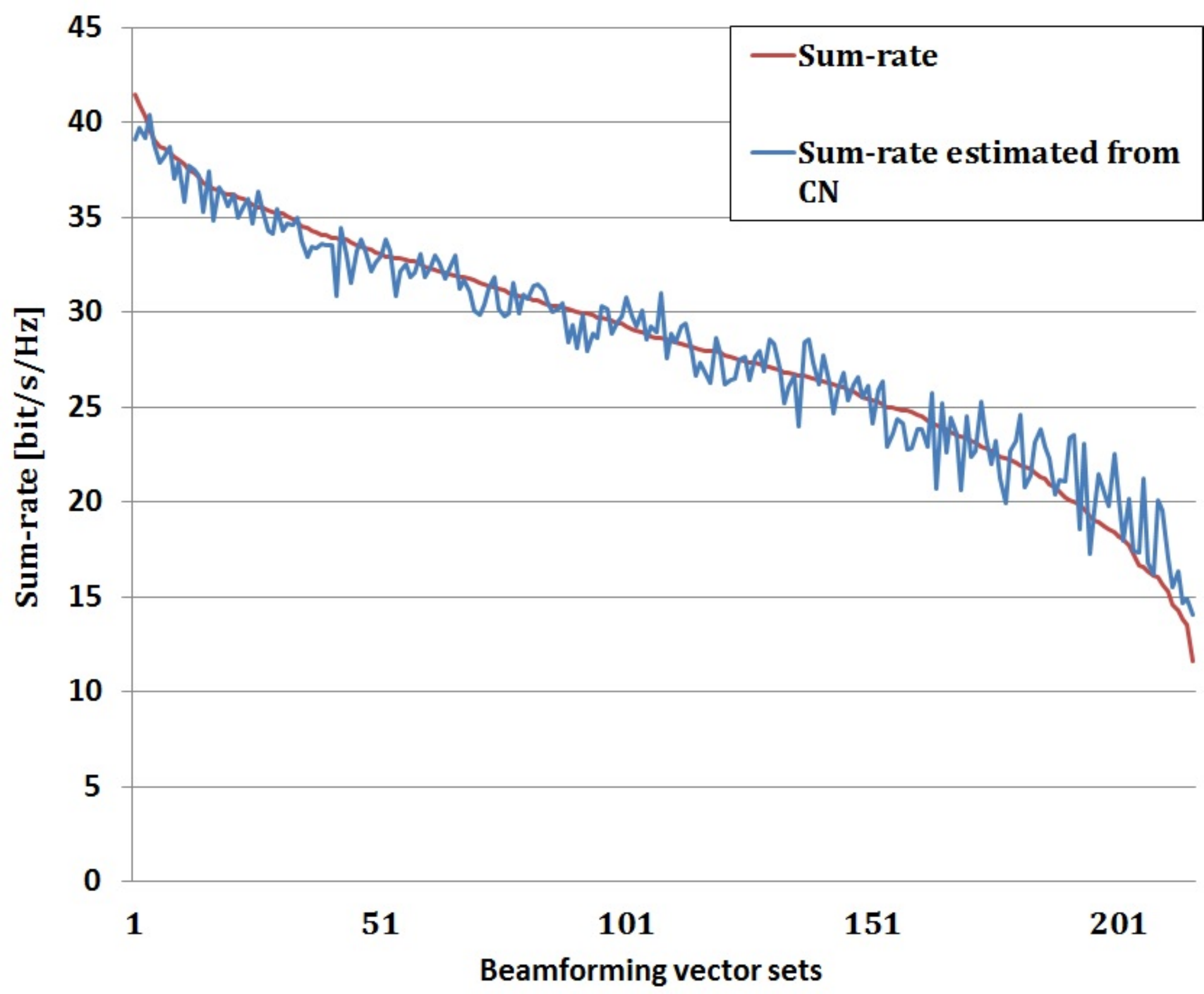}
\caption{Comparison of the directly obtained sum-rate with the estimated sum-rate from CN in the $3 \times 3$ X channel.}
\label{4by3sumrate}
\end{figure}
\subsection{Comparison of Computational Complexity}
The singular values ​​$ \sigma_ {max}, \sigma_ {min} $ of the signal space matrix for CN can be obtained from its SVD.
A null space for each desired signal in the signal space matrix can also be obtained from the SVD of the signal space matrix.
Therefore, the computational complexity of the MinMax SNR based beamformer selection method or the sum-rate maximization based beamformer selection method is equivalent to calculating the SVD approximately $ 3K $ times because there are $3K$ desired signal spaces.
On the other hand, the computational complexity of the CN based beamformer selection method is equivalent to calculating the SVD approximately $ 3 $ times because there are $3$ signal space matrices. Thus, the proposed CN based beamformer selection method reduces its computational complexity compared to the MinMax SNR based beamformer selection method or the sum-rate maximization based beamformer selection method by $1 \over K$.

\vspace{5pt}

\section{Performance Analysis of Expanded Beamformer Sets with CN Based Beamformer Selection}
In this section, the performance of SER and sum-rate of the $K \times 3$ MIMO X channel with CN based beamformer selection method are analyzed.
For simplicity, we consider the case of $K=3$ and assume that each transmitter has the same transmit power constraint $P$.

\subsection{SER by CN Based Beamformer Selection}
For the case of $3 \times 3$ MIMO X channel, the proposed CN based beamformer selection method for SER has the following four steps.
\begin{itemize}
\item For each $\textbf{b}_{j}\in\textbf{B}$, compute CN by $\underset{i\in(1,2,3)}{\operatorname{max}}(\kappa_{i}(\textbf{b}_{j})),\hspace{0.01\textwidth}j=1, 2, \cdots, 6^{3}$ of the normalized signal space matrix obtained from $6^{3}$ beamforming vector sets, where $\kappa_{i}(\textbf{b}_{j})$ is the CN of the $i$-th receiver with the beamforming vector set $\textbf{b}_{j}$ in \textbf{B}.
\item Let $\bar{\textbf{b}}_{j}$ be the sorted beamforming vector in the ascending order of $\underset{i}{\operatorname{max}}(\kappa_{i}(\textbf{b}_{j}))$. 
\item Choose a set of $\bar{\textbf{b}}_{j}$'s with the smallest $u$ CN's.
\item $\textbf{b}^*=\operatorname{arg}\underset{\bar{\textbf{b}}}{\operatorname{max}}\underset{i}{\operatorname{min}}|\textbf{R}^{\dagger}\textbf{H}\textbf{v}|_{i},\hspace{0.03\textwidth} i=1, 2, \cdots, 9$.
\end{itemize}
\begin{figure}[t!]
\centering
\includegraphics[width=0.5\textwidth]{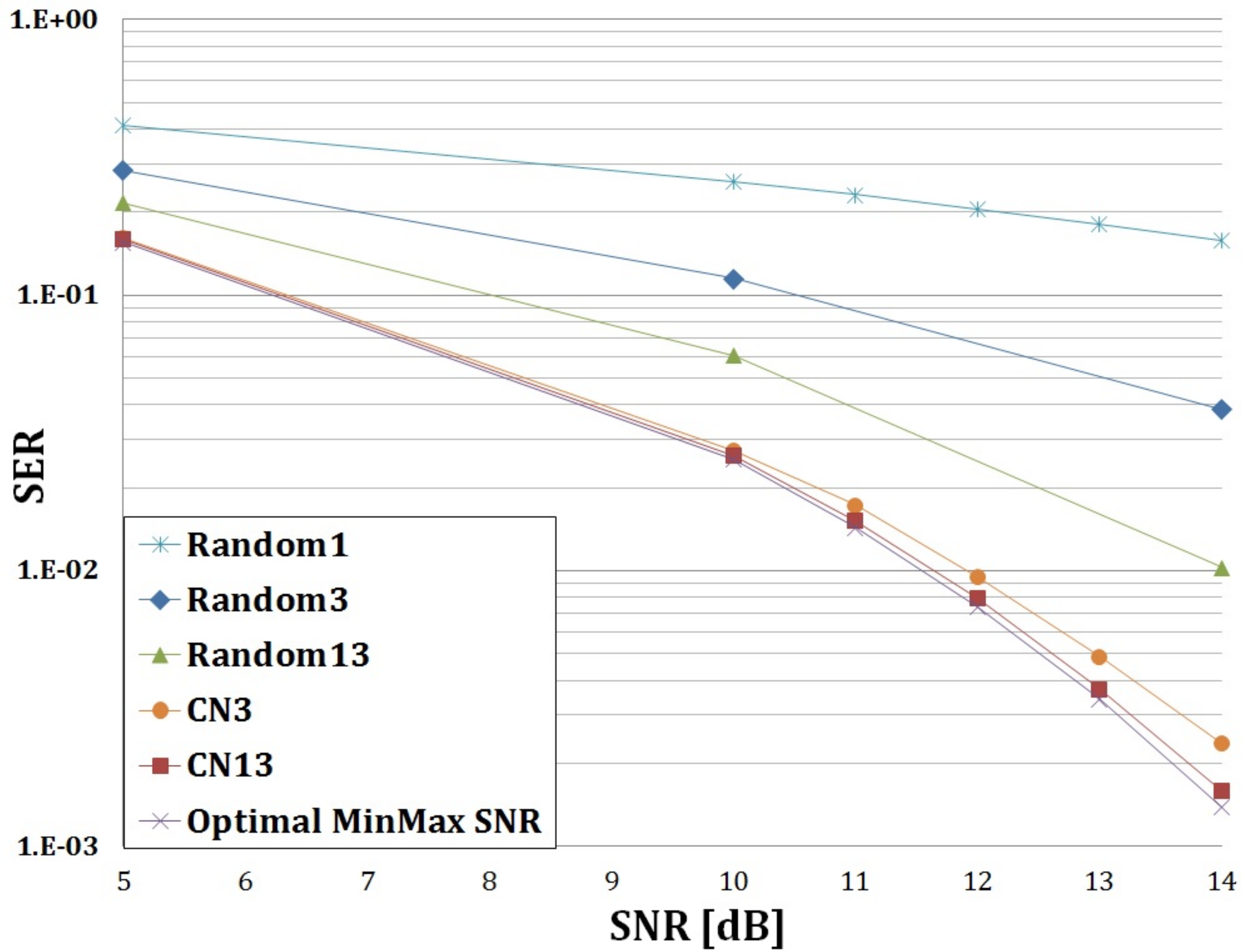}
\caption{SER comparison between the CN based beamformer selection method and the MinMax SNR based beamformer selection method for a Latin square in the case of $3 \times 3$ MIMO X channel.}
\end{figure}

In Fig. 6, Optimal MinMax SNR uses the method to obtain the minimum SNR for each beamforming vector set and select the maximum SNR among them for a Latin square. Random$u$ uses the method to obtain the minimum SNR for each of randomly selected $u$  beamforming vector sets and select the maximum SNR among them. For CN$13$, it has almost similar SER performance to Optimal MinMax SNR. Also, compared to Random$u$, it is very effective to select the beamforming vector sets by CN. Therefore, it can be seen that good SER performance can be achieved by the CN based beamformer selection without obtaining the SNR of all the beamforming vector sets.

\subsection{Sum-Rate by CN Based Beamformer Selection}
For the case of $3 \times 3$ MIMO X channel, the proposed CN based beamformer selection method for the sum-rate is obtained as
\begin{align}
\textbf{b}^*=\operatorname{arg}\underset{\textbf{b}_{j}\in\textbf{B}}{\operatorname{min}}(\sum_{i=1}^{3}(\kappa_{i}(\textbf{b}_{j}))),\hspace{0.01\textwidth}j=1, 2, \cdots, 6^{3}. 
\end{align}
\indent In Fig. 7, Optimal sum-rate uses the method to obtain the sum-rate for each beamforming vector sets and select the maximum sum-rate among them. Random$u$ uses the method to obtain the sum-rate of randomly selected $u$ beamforming vector sets and select the maximum sum-rate among them. For CN$1$, it has almost similar sum-rate performance to Optimal sum-rate. Also, compared to Random$u$, it is very effective to select the beamforming vector sets by CN. Therefore, it can be seen that high sum-rate performance can be achieved by the CN based beamformer selection without obtaining the sum-rate of all the beamforming vector sets.
\begin{figure}[t!]
		\centering
    \includegraphics[width=0.5\textwidth]{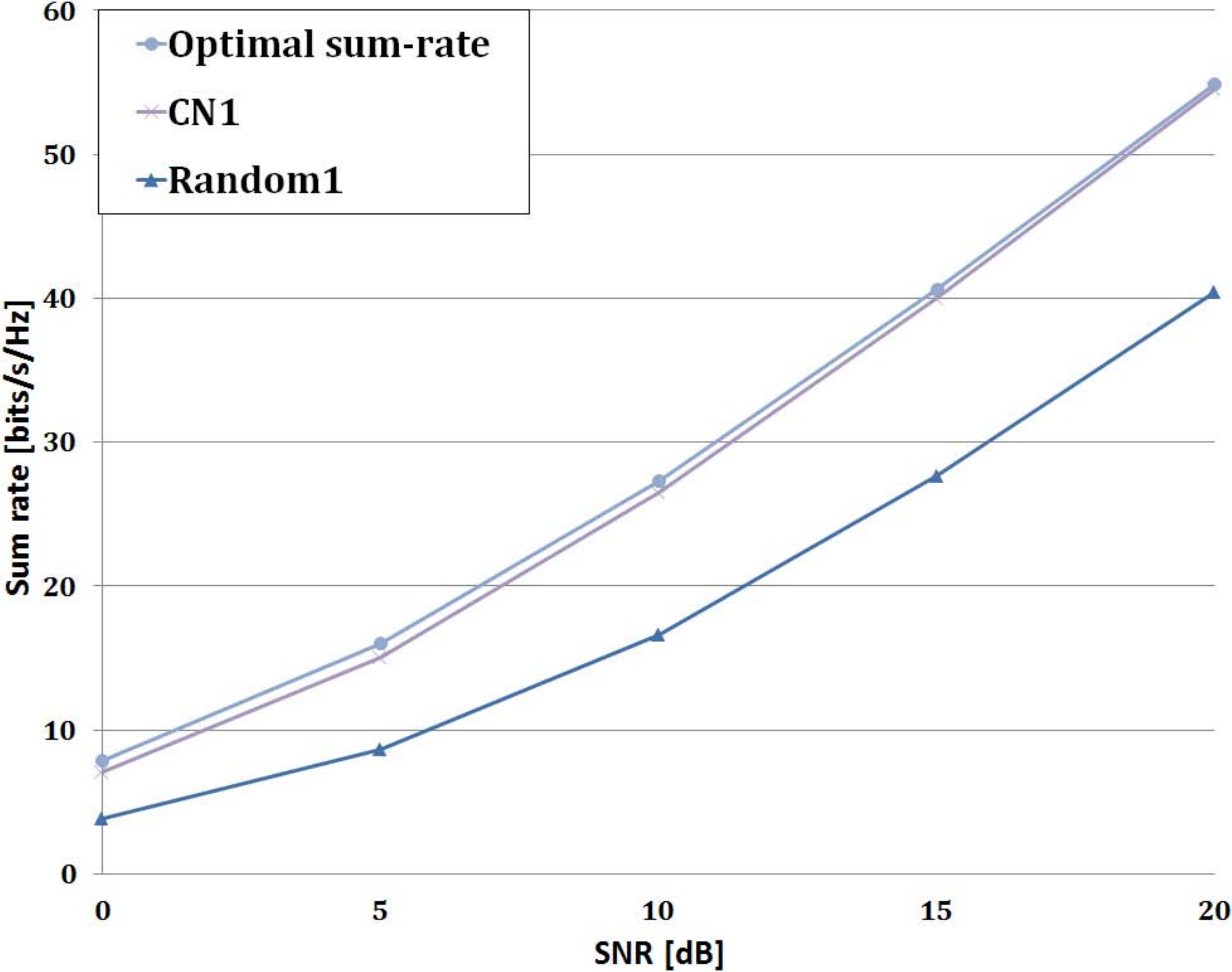}
		\caption{Sum-rate comparison between the CN based beamformer selection method and the sum-rate maximization based beamformer selection method for a Latin square in the case of $3 \times 3$ MIMO X channel.}
	\end{figure}

\subsection{Latin Square Expansion Gain}
The proposed IA scheme for $K \times 3$ MIMO X channel has at least $(K-1)!L(K,K)$ IA schemes by the $K \times K$ Latin squares. Since there are $(2K)^{K}$ beamforming vector sets for each IA scheme, the total number of the expanded beamforming vector sets is at least $(K-1)!L(K,K)(2K)^{K}$. If computational power is sufficient, the selectable range of beamforming vector sets for the proposed IA scheme can be expanded by the Latin square.\\ 
\indent Fig. 8 shows the SER performance for the beamforming vector sets by the Latin square expansion, where Prop$216$ and Prop$432$ stand for a $3 \times 3$ Latin square and two $3 \times 3$ Latin squares, respectively. The Latin square expansion provides more beamforming vector sets and thus SER performance is improved compared to the conventional one. As a further work, it can be researched how to choose the best IA scheme among at least $(K-1)!L(K,K)$ IA schemes.
\begin{figure}[t!]
		\centering
    \includegraphics[width=0.55\textwidth]{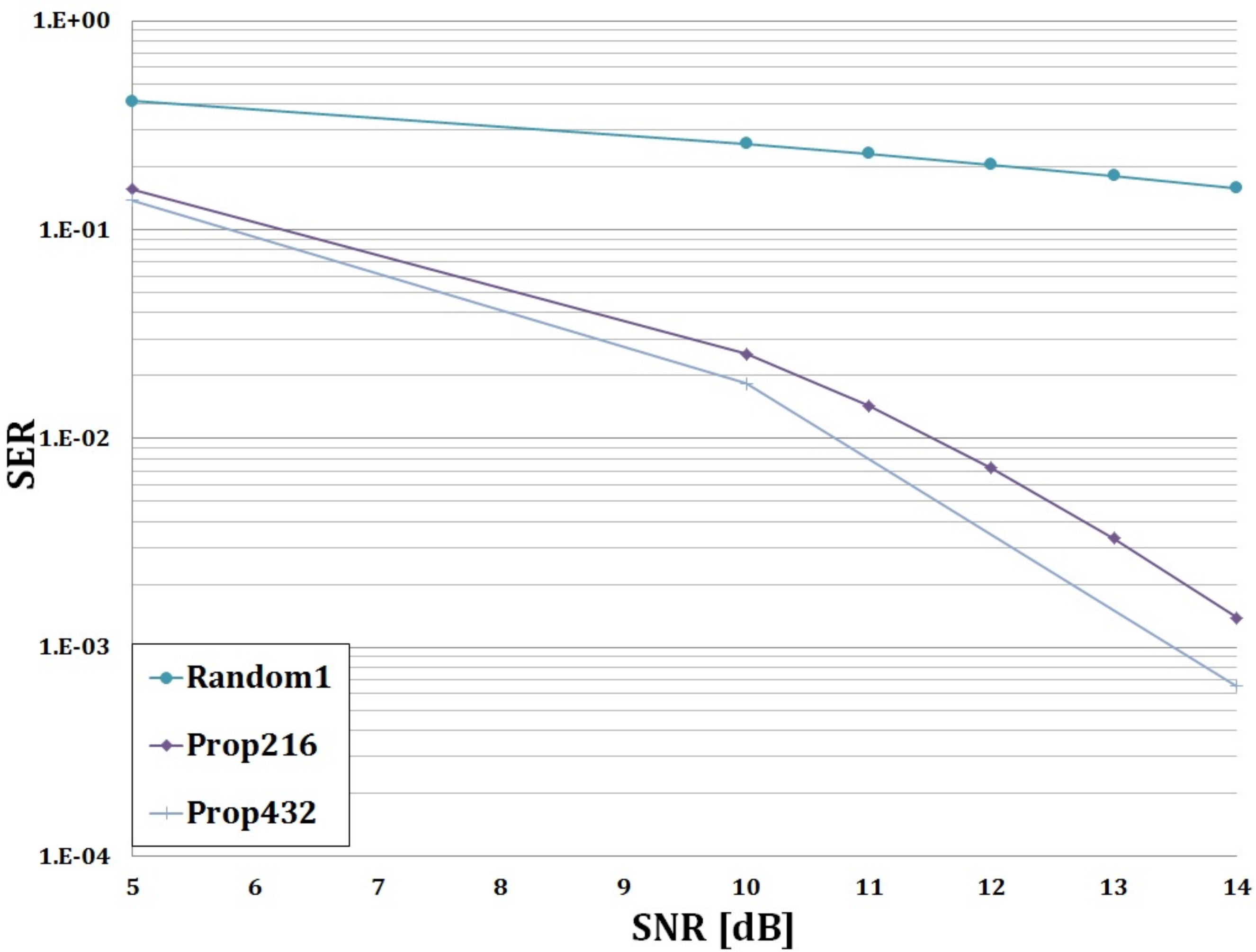}
		\caption{SER comparison of the proposed IA scheme with the Latin square expansion and the conventional scheme for the case of $3 \times 3$ MIMO X channel.}
	\end{figure}

\vspace{5pt}

\section{Conclusions}
In this paper, we proposed an IA scheme using the Latin square and CN based beamformer selection method with low computational complexity in the $K \times 3$ MIMO X channel. It is shown that the proposed IA scheme can be expanded by the $K \times K$ Latin squares. Since the proposed IA scheme has a chain structure, the proposed scheme has a larger beamforming vector sets than the conventional IA scheme. To select the good beamforming vector sets among many beamforming vector sets, the CN based beamformer selection with low computational complexity was proposed. SER and sum-rate performance can be improved by the proposed IA scheme using the Latin squares with the CN based beamformer selection.\\

\vspace{5pt}

\section*{Acknowledgment}

\end{document}